\documentclass[prd,twocolumn,superscriptaddress,amsmath,%
amssymb,nofootinbib]{revtex4-1}

\begin{document}


\title{Cardy formula for charged  black holes with anisotropic scaling}

\author{Mois\'es~Bravo-Gaete}
\email{moisesbravog-at-gmail.com} \affiliation{Facultad de Ciencias
B\'asicas, Universidad Cat\'olica del Maule, Casilla 617, Talca,
Chile}

\author{Sebasti\'an~G\'omez}
\email{sebago-at-inst-mat.utalca.cl} \affiliation{Instituto de
Matem\'atica y F\'isica, Universidad de Talca, Casilla 747 Talca,
Chile}

\author{Mokhtar~Hassa\"ine}
\email{hassaine-at-inst-mat.utalca.cl} \affiliation{Instituto de
Matem\'atica y F\'isica, Universidad de Talca, Casilla 747 Talca,
Chile}

\begin{abstract}
We first observe that for Lifshitz black holes whose only charge is
the mass, the resulting Smarr relation is a direct consequence of
the Lifshitz Cardy formula. From this observation, we propose to
extend the Cardy formula to the case of electrically charged
Lifshitz black holes satisfying as well a Smarr relation. The
expression of our formula depends on the dynamical exponent, the
energy and the charge of the ground state which is played by a
magnetically charged soliton obtained through a double Wick
rotation. The expression also involves a factor multiplying the
chemical potentials which varies in function of the electromagnetic
theory considered. This factor is precisely the one that appears in
the Smarr formula for charged Lifshitz black holes. We test the
validity of this Cardy formula in different situations where
electrically Lifshitz charged black holes satisfying a Smarr
relation are known. We then extend these results to electrically
charged black holes with hyperscaling violation. Finally, an example
in the charged AdS case is also provided.

\end{abstract}

\maketitle

\section{\label{sec:intro}Introduction}
Recently, there has been an important interest in extending the
ideas underlying the standard relativistic AdS/CFT correspondence
\cite{Maldacena:1997re} to physical systems that exhibit a dynamical
scaling near fixed points. These latter are characterized by an
anisotropic invariance encoded by the fact that the space and the
time scale with different weights,
\begin{eqnarray}
t\to\lambda^z\,t,\qquad\qquad \vec{x}\to\,\lambda \vec{x}.
\label{anisyymetry}
\end{eqnarray}
The constant $z$ which is called the dynamical exponent precisely
reflects this anisotropic symmetry. In analogy with the AdS case
$z=1$, the gravity dual metric in $D-$dimensions refereed as the
Lifshitz metric was given in \cite{Kachru},
\begin{equation}
\label{Lifmetric}
ds^2=-\left(\frac{r}{l}\right)^{2z}dt^{2}+\frac{l^{2}}{r^{2}}\,dr^{2}+\frac{r^{2}}{l^{2}}\,\sum_{i=1}^{D-2}dx_{i}^{2},
\end{equation}
and, it is easy to see that the anisotropic transformations
(\ref{anisyymetry}) together with the rule $r\to \lambda^{-1}r$ act
as an isometry for this metric. Nevertheless, in contrast with the
AdS case, Lifshitz spacetimes or their black hole extensions are not
solutions of standard General Relativity, and instead require the
introduction of some source that may be materialized by some extra
fields
\cite{Azeyanagi:2009pr,Correa:2014ika,Ayon-Beato:2015jga,Bravo-Gaete:2013dca}
or/and by considering higher-order gravity theories
\cite{AyonBeato:2009nh,AyonBeato:2010tm,Lu:2012xu,Giacomini:2012hg,Oliva:2012zs}.
The thermodynamical properties  of the Lifshitz black holes, in
spite of their rather unconventional asymptotic behaviors, have been
intensively studied, see e. g.
\cite{Liu:2014dva,Devecioglu:2011yi,Bravo-Gaete:2015xea}. One of the
most appealing property of the Lifshitz black holes whose only
charge is the mass $\Delta$ concerns their entropy ${\cal S}$ which
scales with respect to the temperature $T$ as
\begin{eqnarray}
{\cal S}\propto T^{\frac{D-2}{z}}. \label{scalesS-T}
\end{eqnarray}
As a direct consequence, the Smarr formula \cite{Smarr} takes the
following form \cite{Brenna:2015pqa}
\begin{eqnarray}
\Delta=\frac{D-2}{D+z-2}\,T{\cal S}. \label{smarrF1}
\end{eqnarray}

In three dimensions, this last relation (\ref{smarrF1}) can be
obtained by exploiting the fact that the Lifshitz algebras in two
dimensions with dynamical exponents $z$ and $z^{-1}$ are isomorphic
\cite{Gonzalez:2011nz}. As shown precisely in this last reference,
this isomorphism is translated into a duality between the low and
high temperature regimes, and allows to derive a formula for the
asymptotic growth number of states in three dimensions where the
ground state is played by the soliton obtained through a double Wick
rotation,
\begin{eqnarray}
{\cal S}=2\pi l(z+1)\left[\left(\frac{\Delta_0}{z}\right)^z\,\Delta
\right]^{\frac{1}{z+1}}, \label{cardyuncharged}
\end{eqnarray}
where $-\Delta_0$ corresponds to the mass of the soliton. In the
isotropic case $z=1$, this expression becomes the standard Cardy
formula. Note also that the validity of Eq. (\ref{cardyuncharged})
has been checked in the case of the Lifshitz black hole solution
with $z=3$ of new massive gravity \cite {AyonBeato:2009nh} (see Ref.
\cite{Gonzalez:2011nz}), and also in presence of a source given by a
nonminimal scalar field  for the same gravity theory
\cite{Ayon-Beato:2015jga}. The first law $d\Delta=Td{\cal S}$
applied to the relation (\ref{cardyuncharged}) will then imply that
the mass can be expressed as
\begin{eqnarray}
\Delta=\frac{\Delta_0}{z}\left(2\pi l\,T\right)^{1+\frac{1}{z}},
\label{massuncharged}
\end{eqnarray}
and combining together the two expressions
(\ref{cardyuncharged}-\ref{massuncharged}), one easily obtains the
Smarr formula (\ref{smarrF1}) for $D=3$. Hence from this simple
exercise, we have highlighted a certain correlation between the
Smarr formula and the generalized Cardy formula in three dimensions.

The main aim of this paper is to extend the formula
(\ref{cardyuncharged}) to the charged case. In doing so, we will
inspire ourselves from the fact that the Smarr formula in the case
of charged solutions must be a consequence of the Cardy formula as
it occurs in the neutral case. This problem has a certain interest
since electrically charged Lifshitz black holes have also be found
in the current literature, see e. g.
\cite{Taylor,DanielssonI,Pang:2009pd,Alvarez:2014pra,Zangeneh:2015uwa,Dehghani:2013mba}.
Such examples occur for example in the case of Einstein gravity with
a source given by a Proca-Maxwell action \cite{Pang:2009pd} or in
presence of $N-$Abelians $U(1)$ fields with a dilaton
\cite{Tarrio:2011de} as well as in the case of nonlinear
electrodynamics \cite{Alvarez:2014pra,Zangeneh:2015uwa}. In all
these examples, a Smarr formula generalizing the expression
(\ref{smarrF1}) can be derived  and, is generically written as
\cite{Brenna:2015pqa}
\begin{eqnarray}
\Delta=\frac{D-2}{D+z-2}\,T{\cal S}+\alpha\, \Phi_e {\cal Q}_e,
\label{smarrF1charged}
\end{eqnarray}
where $\Phi_e$ is the electric potential and ${\cal Q}_e$ the
electric charge. In this relation, the value of the constant
$\alpha$ varies in function of the electromagnetic Lagrangian
considered. From now, it is important to emphasize the non-universal
character of the Smarr formula in the charged case reflected by the
presence of the constant $\alpha$. In other words, this means that
the constant $\alpha$ does not depend only on the dynamical exponent
$z$ and the dimension $D$ but also depends on the theory considered
as we will see in the different examples listed below.

In this paper, we will show that for electrically charged Lifshitz
black holes  satisfying a Smarr relation of the form
(\ref{smarrF1charged}) in three dimensions, the Cardy formula
(\ref{cardyuncharged}) becomes
\begin{eqnarray}
{\cal S}=2\pi l(z+1)\Big(\vert \Delta_0\,z^{-1}+\alpha\, \Phi_m
{\cal Q}_m \vert^z\,\vert\Delta-\alpha\,
\Phi_e {\cal Q}_e \vert\Big)^{\frac{1}{z+1}},\nonumber\\
\label{cardycharged}
\end{eqnarray}
where $\Phi_m$ (resp.  ${\cal Q}_m$) denotes the magnetic potential
(resp. the magnetic charge) of the magnetically charged soliton
obtained from the electric solution by means of a double Wick
rotation. Since the Wick rotation switches the role of the time
coordinate $t$ with the angular coordinate $x=\varphi$, the field
strengths of the resulting magnetically charged soliton will be in
general complex with a magnetic charge and potential both purely
imaginary. Nevertheless, this will not be dramatic since in the
proposed formula (\ref{cardycharged}), it only appears their product
which is always real. The Wick rotation is also responsible of the
apparent discrepancy of the sign appearing in front of the constant
$\alpha$ accompanying  the magnetic  and electric parts in
(\ref{cardycharged}).

In what follows, we will test the validity of the formula
(\ref{cardycharged}) in different theories where charged Lifshitz
black holes satisfying a Smarr formula of the form
(\ref{smarrF1charged}) are known. In each case, we will derive the
corresponding magnetically charged soliton and compute their mass
through the quasilocal method given in
\cite{Kim:2013zha,Gim:2014nba} as well as their magnetic charge. We
will then extend these results to the case of charged hyperscaling
violation black holes. Finally, the last section will be dedicated
to some comments regarding the isotropic AdS case $z=1$.

\section{Charged Lifshitz black hole and soliton solutions}

In all the examples given below, the Lagrangian ${\cal L}$ will
involve a gravity part encoded by the metric $g$ as well as
different Abelian fields denoted generically by $A_{(i)\mu}$ and
eventually a scalar field $\phi$ with its standard kinetic term
$\partial_{\mu}\phi\partial^{\mu}\phi$,
\begin{eqnarray}
{\cal L}={\cal L}\left(g, A_{(i)\mu}, \phi\right). \label{Lg}
\end{eqnarray}
The corresponding action will be given by
\begin{eqnarray}
S[g,\phi, A_{(i)\mu}]=\int d^3x\,\sqrt{-g} \,{\cal L}.
\end{eqnarray}

The mass of the charged black hole and soliton will be computed
through the quasilocal method described in Refs.
\cite{Kim:2013zha,Gim:2014nba} where the charge $\Delta$ which
corresponds to the mass is given by
\begin{equation}\label{eq:5}
\Delta(\xi)=\int_{\cal B}\!dx_{\mu\nu}
\Big(\delta{K}^{\mu\nu}(\xi)-2\xi^{[\mu}\!\!
\int^1_0\!\!ds~\Theta^{\nu]}(\xi|s)\Big).
\end{equation}
Here $\delta{K}^{\mu\nu}(\xi)\equiv
{K}^{\mu\nu}_{s=1}(\xi)-K^{\mu\nu}_{s=0}(\xi)$ denotes the
difference of the Noether potential between the interpolated
solutions, $dx_{\mu\nu}$ represents the integration over the
co-dimension two boundary ${\cal B}$, $\xi^t=(1,0,0)$ is the
timelike Killing vector field and $\Theta^{\nu}$ represents the
surface term. In the case of a Lagrangian given by (\ref{Lg}), the
involved quantities are given by
\begin{eqnarray}
\Theta^\mu &=&2\sqrt{-g}\Big[P^{\mu(\alpha
\beta)\gamma}\nabla_\gamma\delta g_{\alpha\beta} -\delta
g_{\alpha\beta}\nabla_\gamma P^{\mu(\alpha\beta)\gamma}
\nonumber\\
&+&\frac{1}{2}\,\displaystyle{\sum_{i}}\left(\frac{\partial
\mathcal{L}}{\partial \left(\partial_{\mu}A_{(i)\nu}\right)}\delta
A_{(i)\nu}\right) +\frac{1}{2}\,\frac{\partial \mathcal{L}}
{\partial \big(\partial_{\mu}\,\phi\big)}\delta \phi\Big],
\label{eq:theta}\\ \nonumber\\
K^{\mu\nu}
&=&\sqrt{-g}\,\left[2P^{\mu\nu\rho\sigma}\nabla_\rho\xi_\sigma
-4\xi_\sigma\nabla_\rho
P^{\mu\nu\rho\sigma}\right.\nonumber\\
&&\left.-\displaystyle{\sum_{i}}\,\frac{\partial
\mathcal{L}}{\partial \left(\partial_{\mu}A_{(i)\nu}\right)}
\xi^{\sigma} A_{(i)\sigma}\right] \label{eq:K},
\end{eqnarray}
where $P^{\mu\nu\rho\sigma}=\frac{\partial {\cal L}}{\partial
R_{\mu\nu\rho\sigma}}$ with $R_{\mu\nu\rho\sigma}$ being the Riemann
tensor.

The black hole metric will be parameterized by the following line
element
\begin{equation}\label{lifbhsoln2}
ds^2 = -\frac{r^{2z}}{l^{2z}}f(r)dt^2
 + \frac{l^2}{r^2f(r)}{dr^2} + \frac{r^2}{l^2}d{\varphi}^2,
\end{equation}
and the Ansatz for the gauge fields and eventually the scalar field
read
\begin{eqnarray}
A_{(i)\mu}dx^{\mu}=A_{(i)t}(r)dt,\qquad \phi=\phi(r).
\label{ElecAnsatz}
\end{eqnarray}
The Euclidean version of (\ref{lifbhsoln2}) obtained by means of the
transformation $t=i\tau$ requires the Euclidean time to be periodic
with period $\beta=T^{-1}$ in order to avoid conical singularity
while the angle keeps identified as $0\leq\varphi<2{\pi}l$. Under
the Euclidean diffeomorphism defined by
\begin{equation}\label{diffeo}
(\tau,r,\varphi)\mapsto\left(\bar{\tau}=\left(\frac{2\pi
l}{\beta}\right)^{\frac{1}{z}}\varphi, \bar{r}=\frac{\beta}{2\pi
z}\left(\frac{r}{l}\right)^z, \bar{\varphi}=\frac{2\pi
l}{\beta}\tau\right),
\end{equation}
the Euclidean Lifshitz black hole is diffeomorphic to another
asymptotically Lifshitz solution with dynamical exponent $z^{-1}$,
scale $lz^{-1}$ and inverse temperature
\begin{equation}
\bar{\beta}=\left(2\pi l\right)^{1+\frac{1}{z}}\beta^{-\frac{1}{z}},
\end{equation}
and finally the Lorentzian soliton will be obtained from
$\bar{\tau}=i\bar{t}$ yielding
\begin{eqnarray}
\label{lifsolsoln1}
ds^2=-\left(\frac{z\bar{r}}{l}\right)^{\frac{2}{z}}d\bar{t}^2
+\frac{l^2}{z^2\bar{r}^2 h(\bar{r})}d\bar{r}^2
+\frac{z^2\bar{r}^2}{l^2}h(\bar{r})d\bar{\varphi}^2.\nonumber\\
\end{eqnarray}
As mentioned before, this double Wick rotation will be responsible
of the fact that the field strengths of the corresponding soliton
will be purely imaginary. Note that in the case of scalar field
which depends only on the radial coordinate, this double Wick
rotation does not yield to a complex scalar field for the soliton
solution \cite{Ayon-Beato:2015jga}. We may also emphasize that the
set of parameters as well as the range of admissible values of the
dynamical exponent $z$ are the same for the electrically charged
black hole and for the magnetically charged soliton.

Also in order to simplify the expressions, the volume of the
one-dimensional sphere is denoted by $\Omega_1$ with
$$
\Omega_1=2\pi l.
$$

We are now in position to check the validity of the expression
(\ref{cardycharged}) in different contexts presented below.

\subsection{Case of Einstein gravity with two Abelian fields and a dilaton}
We first analyze the case of Einstein gravity with two Abelian
fields and a dilaton for which the Lagrangian reads
\begin{eqnarray}
{\cal
L}=\frac{1}{2\kappa}\left(R-2\Lambda-\frac{1}{2}\partial_{\mu}\phi\partial^{\mu}\phi-\frac{1}{4}\sum_{i=1}^2e^{\lambda_i\phi}F_{(i)}^2\right),
\label{Lagdila}
\end{eqnarray}
with $F_{(i)}^2=F_{(i)\mu\nu}F_{(i)}^{\,\,\mu\nu}$ for $i=1, 2$.

For an Ansatz of the form (\ref{lifbhsoln2}-\ref{ElecAnsatz}), the
solution given in \cite{Tarrio:2011de} reads
\begin{subequations}
\label{sol2u1}
\begin{eqnarray}
&&f(r)=1-m\left(\frac{r_{h}}{r}\right)^{z+1}+(m-1)\left(\frac{r_{h}}{r}\right)^{2z},\\
\label{sol2u1a}
&& F_{(1)rt}=\sqrt{2\,(z^2-1)}\,\mu^{\sqrt{\frac{1}{2(z-1)}}}\,\left(\frac{r}{l}\right)^z,\\
&&F_{(2)rt}=\sqrt {2(m-1)(z-1)}{\mu}^{-\sqrt {\frac{z-1}{2}}}
\left(\frac{r_h}{l}\right)^{z}\,r^{-z},\\
\label{sol2u1b} && e^{\phi}=\mu \,r^{\sqrt{2(z-1)}},
\end{eqnarray}
\end{subequations}
where $m$ and $\mu$ are two integration constants and $r_{h}$ stands
for the location of the horizon. Note that we have opted for this
parametrization of the solution for latter convenience but the
expressions (\ref{sol2u1}) are equivalent to those given in
\cite{Tarrio:2011de} after some redefinitions of the constants. This
solution is defined provided that the parameters are fixed as
follows
\begin{eqnarray}
\Lambda=-\frac{z(z+1)}{2l^2},\, \lambda_1=-\sqrt{\frac{2}{z-1}},\,
\lambda_2=\sqrt{2(z-1)}, \label{param1}
\end{eqnarray}
while the range for the admissible values of the dynamical exponent
is $z>1$.

In this case, the Wald entropy together with the Hawking temperature
read
\begin{subequations}
\begin{eqnarray}
\label{Wald1}
\mathcal{S}_{\mathrm{W}}&=&\frac {2\,\pi\,\Omega_{1}}{\kappa} \left(\frac{r_h}{l}\right),\\
T&=&\frac{1}{4 \pi l}\,\left[ 2\,z+ \left( 1-z \right) m\right]
\left(\frac{r_{h}}{l}\right)^{z}=\frac{\sigma}{l}
\left(\frac{r_{h}}{l}\right)^{z},
\end{eqnarray}
\end{subequations}
where we have defined
\begin{eqnarray}\label{expressofA}
\sigma=\frac{1}{4\,\pi}\left[2\,z+ \left( 1-z \right) m\right].
\end{eqnarray}
On the other hand, the electric charge and electric potential read
respectively
\begin{eqnarray}
{\cal Q}_{e}= \frac {\sqrt {2(m-1)(z-1)}\, {\mu}^{\frac{1}{2} \sqrt
{2\,(z-1)}}\,\Omega_{1}}{2\,\kappa\, {l}^{2}}\,{r_h}^{z},
\end{eqnarray}
and
\begin{eqnarray}
\Phi_{e}=-A_{(2) t}(r_{h})=\frac {\sqrt
{2\,(m-1)}\,{\mu}^{-\frac{1}{2} \sqrt {2\,(z-1)}}}{\sqrt
{z-1}\,{l}^{z}}\,{r_h}.
\end{eqnarray}

Introducing a one-parameter family of locally equivalent solutions,
the variation of the Noether potential and the surface term
(\ref{eq:theta}-\ref{eq:K}) are given by
\begin{align}
\delta K^{r t}&= -\frac
{(z-1)\,m}{2\,\kappa\,l}\left(\frac{r_{h}}{l}\right)^{z+1}+\frac {
\left( m-1 \right)\,r_h^{2\,z}}{\kappa\,l^{z+2}}\,{r}^{-z+1}
,\nonumber\\
\int_{0}^{1}\!\!\!\!d s\Theta^{r}&=\frac
{m\,z}{2\,\kappa\,l}\left(\frac{r_{h}}{l}\right)^{z+1}-\frac {
\left( m-1
\right)\,r_h^{2\,z}}{\kappa\,l^{z+2}}\,{r}^{-z+1}.\nonumber
\end{align}

From these expressions, we obtain the mass of the Lifshitz black
hole to be
\begin{equation}\label{deltap1}
\Delta=\frac{m \, \Omega_{1}}{2\,\kappa\, l}
\left(\frac{r_{h}}{l}\right)^{z+1},
\end{equation}
and we easily check  that the first law holds
\begin{eqnarray}\label{firstlaw}
d \Delta=Td\mathcal{S}_{\mathrm{W}}+\Phi_e d{\cal Q}_e.
\end{eqnarray}
The Smarr formula turns to be
\begin{eqnarray}
\Delta=\frac{1}{z+1}\left(T\mathcal{S}_{\mathrm{W}}+z\Phi_e {\cal
Q}_e\right), \label{smarrDilaton}
\end{eqnarray}
and corresponds to the expression (\ref{smarrF1charged}) with
$\alpha=\frac{z}{z+1}$.

The metric function of the corresponding solitonic spacetime
(\ref{lifsolsoln1}) is given by
\begin{eqnarray}
h(\bar{r})=1-\frac{m}{(2 \pi \sigma)^{\frac{z+1}{z}}}
\left(\frac{l}{z \bar{r}}\right)^{\frac{z+1}{z}}+\frac{(m-1)}{(2 \pi
\sigma)^2}\left(\frac{l}{z \bar{r}}\right)^2,
\end{eqnarray}
where $\sigma$ is defined in (\ref{expressofA}), and the Abelian
gauge fields and the dilaton read
\begin{subequations}
\begin{eqnarray}
&& F_{(1)\bar{r}\bar{\varphi}}=i\frac{\sqrt{2\,(z^2-1)}\,\mu^{\sqrt{\frac{1}{2(z-1)}}}}{l}\,\,\bar{r}^{\frac{1}{z}}, \label{F1rvarphip1}\\
&&F_{(2)\bar{r} \bar{\varphi}}=i\frac{\sqrt {2(m-1)(z-1)}\,{\mu}^{-\sqrt {\frac{z-1}{2}}}}{2 \pi \sigma z}\,\bar{r}^{\frac{1-2z}{z}},\\
&& e^{\phi}=\mu \,\bar{r}^{\frac{\sqrt{2(z-1)}}{z}} \label{phip1}.
\end{eqnarray}
\end{subequations}
As before,  the variation of the Noether potential and the surface
term read
\begin{align}
\delta K^{\bar{r} \bar{t}}&=\frac{1}{2\kappa} \left[{\frac
{(m-1)}{2\,{\pi }^{2}{\sigma}^{2}\,z\,\bar{r}}} \,\left( {\frac
{zr}{l}} \right) ^{\frac{1}{z}}-\frac{2\,m} {  \left( 2 \pi \sigma
\right) ^{{\frac {z+1}{z}}}{l}}\right],
\nonumber\\
\int_{0}^{1}\!\!\!\!d s\Theta^{\bar{r}}&=\frac{1}{2\kappa} \left[
-{\frac {(m-1)}{2\,{\pi }^{2}{\sigma}^{2}\,z\,\bar{r}}} \,\left(
{\frac {z\bar{r}}{l}} \right) ^{\frac{1}{z}} +\frac{m (z+2)} {
\left( 2 \pi \sigma \right) ^{{\frac {z+1}{z}}}{l}}\right],\nonumber
\end{align}
yielding to
\begin{eqnarray}\label{minussolmass}
\Delta_{0}=\frac{z\,m\,\Omega_{1}} { 2 \kappa l \left( 2 \pi
\sigma\right) ^{{\frac {z+1}{z}}}}.
\end{eqnarray}
Finally, the magnetic charge and potential are respectively
expressed as
\begin{eqnarray}
{\cal Q}_{m}= i\frac {\sqrt {2(m-1)(z-1)}\, {\mu}^{\frac{1}{2} \sqrt
{2\,(z-1)}}\,\Omega_{1}}{4\,\pi
\sigma\,z\,\kappa}\,\left(\frac{z}{l}\right)^{\frac{1}{z}},
\end{eqnarray}
and
\begin{eqnarray}
\Phi_{m}=i\frac {\sqrt {2\,(m-1)}\,z\,{\mu}^{-\frac{1}{2} \sqrt
{2\,(z-1)}}}{\sqrt {z-1}\,{l}}\,\left(\frac{l}{2 \pi \sigma
z}\right)^{\frac{1}{z}}.
\end{eqnarray}

It is then easy to verify that the formula (\ref{cardycharged}) with
the parameter $\alpha=\frac{z}{z+1}$ correctly fits with the
expression of the Wald entropy given by Eq. (\ref{Wald1}).

\subsection{Case of Einstein gravity with a nonlinear electrodynamics}
In Ref. \cite{Zangeneh:2015uwa}, the authors consider a slightly
generalization of the previous Lagrangian (\ref{Lagdila}) by
introducing a nonlinear term as
\begin{eqnarray}
{\cal
L}=\frac{1}{2\kappa}\Big[&&R-2\Lambda-\frac{1}{2}\partial_{\mu}\phi\partial^{\mu}\phi-\frac{1}{4}e^{\lambda_1\phi}F_{(1)}^2\nonumber\\
&&+\left(-\frac{1}{4}e^{\lambda_2\phi}F_{(2)}^2\right)^p\Big].
\label{Lagdilanonli}
\end{eqnarray}
We have made some redefinitions of the fields and parameters in the
original action \cite{Zangeneh:2015uwa} such that the Lagrangian
(\ref{Lagdilanonli}) reduces to (\ref{Lagdila}) in the linear
limiting case $p=1$. Note that such nonlinear generalization of the
Maxwell action has been currently studied, see e. g.
\cite{nonlinMax}.

For an Ansatz of the form (\ref{lifbhsoln2}-\ref{ElecAnsatz}), the
metric function given in \cite{Zangeneh:2015uwa} after some
redefinitions of the constants reads
\begin{eqnarray}\label{metricfunctionp}
&&f(r)=1-m\left(\frac{r_{h}}{r}\right)^{z+1}+(m-1)\left(\frac{r_{h}}{r}\right)^{2
z+\Gamma},
\end{eqnarray}
where the constant $\Gamma$ is defined as
\begin{eqnarray}\label{Gamma}
\Gamma=-\frac{2(p-1)}{2\,p-1}.
\end{eqnarray}
For this solution, the uncharged Abelian field $F_{(1)rt}$, the
dilaton, the cosmological constant and the coupling constants are
given by the same expressions than in the linear case, see
(\ref{sol2u1a}), (\ref{sol2u1b}) and (\ref{param1}). The only
changes are concerned with the charged Abelian gauge field
$F_{(2)rt}$ and the coupling constant $\lambda_2$ which now take the
following forms
\begin{eqnarray*}
&&F_{(2) rt}=\frac{\sqrt {2}\, \big(r_h\big)^{{\frac { \left( 2\,z-1
\right) p-z+1}{
 \left( 2\,p-1 \right) p}}}\,\Sigma^{\frac{1}{2p}}\,}{{l}^{{\frac {1+ \left( z-1 \right) p}{p}}}\,{\mu}^
{{\frac { \left( z-1 \right) \left( 2\,p-1 \right) }{\sqrt {2(
z-1)}p}}}\,{r}^{{\frac { 2\left( z-1 \right) p-z+2} {2\,p-1}}}},\\
&&{\lambda_2}={\frac {2\,[ 2\left( z-1 \right) p-z+1]} {p\,\sqrt
{2(z-1)}}},
\end{eqnarray*}
where for simplicity we have defined
\begin{eqnarray}\label{Sigmadilp}
\Sigma={\frac { \left( m-1 \right)  \left[  2\left( z-2 \right)
p-z+3 \right] }{ \left( 2\,p-1 \right) ^{2}}}.
\end{eqnarray}

The expression of the entropy as well is unchanged and given by
\begin{eqnarray}
\mathcal{S}_{\mathrm{W}}&=&\frac {2\,\pi\,\Omega_{1}}{\kappa}
\left(\frac{r_h}{l}\right), \label{Wald1nonlinear}
\end{eqnarray}
where $r_{h}$ is now the location of the horizon for the metric
function (\ref{metricfunctionp}). The Hawking temperature for this
configuration reads
$$T=\frac{1}{l}\,\left[ \sigma+{\frac
{ \left( p-1 \right)  \left( m-1 \right) } {2\,\pi \, \left( 2\,p-1
\right) }} \right]  \left( {\frac {{r_h}}{l}}
 \right) ^{z}
 =\frac{\bar{\sigma}}{l}\, \left( {\frac {{r_h}}{l}}
 \right) ^{z},
$$
where we have defined
\begin{eqnarray}\label{barsigma}
\bar{\sigma}=\sigma+{\frac { \left( p-1 \right)  \left( m-1 \right)
} {2\,\pi \, \left( 2\,p-1 \right) }},
\end{eqnarray}
with $\sigma$ given by (\ref{expressofA}). On the other hand, the
electric potential together with the electric charge read
\begin{align}
\Phi_{e}&=\frac{\sqrt {2}\left( 2\,p-1 \right)
 \big(r_h\big)^{-{\frac {[ \left( z-2
\right) p-z+1]}{p}}}\,{\Sigma}^{\frac{1}{2 p}} }{{l}^{{\frac {1+
\left( z-1 \right) p}{p}}} \left[ 2\left( z-2 \right) p-z+3 \right]
{\mu}^{{\frac { \left( z-1 \right)  \left( 2\, p-1 \right) }{p \sqrt
{2(z-1)}}}} },
\end{align}
\begin{align}
{\cal{Q}}_{e}&=\frac{\sqrt {2}\,p\,{\mu}^{{\frac { \left( z-1
\right) \left( 2\,p-1 \right) }{p \sqrt {2(z-1)}}}}{\Sigma}^{{\frac
{2 p-1}{2 p}}} \big(r_h\big)^ {{\frac { \left( 2\,z-1 \right)
p-z+1}{p}} }\Omega_{1} }{2\,\kappa\,l^{\frac{3p-1}{p}}},
\end{align}
with $\Sigma$ given by the expression (\ref{Sigmadilp}).

Let us now compute the mass of this solution through the quasilocal
formalism. For the timelike Killing vector $\xi^{t}=(1,0,0),$ and
after some tedious but straightforward computations, the surface
term together with the variation of the Noether potential
(\ref{eq:theta}-\ref{eq:K}) are given by
\begin{align}
\int_{0}^{1}\!\!\!\!d s\Theta^{r}&=- \frac{\left( m-1 \right)
{\big(r_h\big)}^{{\frac { 2 \left( 2\,z-1 \right) p-2\,z
+2}{2\,p-1}}}{r}^{-{\frac {[ 2\left( z-2 \right)
p-z+3]}{2\,p-1}}}}{{ \kappa}  {l}^{z+2}}\nonumber\\
&+\frac{zm}{2\, \kappa\, l} \left( {\frac {r_h}{l}} \right) ^{1+z}
,\nonumber\\
\delta K^{r t}&=\frac{\left( m-1 \right) {\big(r_h\big)}^{{\frac { 2
\left( 2\,z-1 \right) p-2\,z +2}{2\,p-1}}}{r}^{-{\frac {[ 2\left(
z-2 \right)
p-z+3]}{2\,p-1}}}}{{ \kappa}  {l}^{z+2}}\nonumber\\
&-\frac{(z-1)\,m}{2\, \kappa\, l} \left( {\frac {r_h}{l}} \right)
^{1+z}.\nonumber
\end{align}
This implies that the mass of the Lifshitz black hole is given by
$$\Delta=\frac{m\,\Omega_{1}}{2 \kappa l}
\left(\frac{r_{h}}{l}\right)^{z+1},$$ and it is simple to verify
that the first law (\ref{firstlaw}) still holds. Additionally, the
Smarr formula turns to be
\begin{eqnarray*}
\Delta=\frac{1}{z+1}\, T\mathcal{S}_{\mathrm{W}}+\left[{\frac
{z}{1+z}}+{\frac { \left( z-1 \right)  \left( p-1 \right) }{
 \left( 1+z \right) p}}\right]\Phi_e {\cal
Q}_e,
\end{eqnarray*}
and corresponds to the expression (\ref{smarrF1charged}) with
\begin{equation}\label{alphap}
\alpha=\frac{z(2p-1)-(p-1)}{p(z+1)}.
\end{equation}

As in the linear case, operating the same diffeomorphism
(\ref{diffeo}), the metric function of the corresponding soliton
reads
\begin{eqnarray}
h(\bar{r})=1-\frac{m}{(2 \pi \bar{\sigma})^{\frac{z+1}{z}}}
\left(\frac{l}{z \bar{r}}\right)^{\frac{z+1}{z}} +\frac{(m-1)}{(2
\pi \bar{\sigma})^{2+\frac{\Gamma}{z}}}\left(\frac{l}{z
\bar{r}}\right)^{2+ \frac{\Gamma}{z}},\nonumber \\
\end{eqnarray}
where $\bar{\sigma}$ is defined in (\ref{barsigma}) and $\Gamma$ is
given in (\ref{Gamma}). As before, the Abelian gauge field $F_{(1)
\bar{r} \bar{\varphi}}$ together with the dilaton are given by
(\ref{F1rvarphip1}) and (\ref{phip1}) respectively, while $F_{(2)
\bar{r} \bar{\varphi}}$ yields
\begin{eqnarray*}
&&F_{(2)\bar{r} \bar{\varphi}}= i\,\sqrt {2}\,\left( {\frac {z}{l}}
\right) ^{{\frac {p-1}{ \left( 2\,p-1 \right) z p}}}\frac{
{\bar{\Sigma}}^{\frac{1}{2\,{p}}}\,}{{\mu}^{{ \frac { \left( z-1
\right) \left( 2\,p-1 \right) }{p\,\sqrt {2\,(z-1)}}} }{\bar{r}}
^{{\frac { 4 \left( z-1\right) p+3-2\,z}{ \left( 2\,p-1 \right)
z}}}},
\end{eqnarray*}
where
$$\bar{\Sigma}={\frac { \left[  2\left( z-2 \right) p+3-z \right]  \left( m-1
 \right) }{ \left( 2\,\pi \,\bar{\sigma} \right) ^{2+\frac{\Gamma}{z}}{z}^{2} \left( 2\,p-1 \right)
 ^{2}}}.
$$
For the same timelike Killing vector, the variation of the Noether
potential and the surface term yield
\begin{equation*}
\delta K^{\bar{r} \bar{t}}= {\frac {(m-1)}{\kappa (2\,{\pi
}{\bar{\sigma}})^{2+\frac{\Gamma}{z}}\,z\,\bar{r}}} \,\left( {\frac
{z\bar{r}}{l}} \right) ^{\frac{4\,p-3}{z (2\,p-1)}} - \frac{m}
{\kappa  \left( 2 \pi \bar{\sigma} \right) ^ {{\frac {z+1}{z}}}{l}}
,
\end{equation*}
\begin{equation*}
 \int_{0}^{1} d s\Theta^{\bar{r}}=
 -{\frac
{(m-1)}{\kappa (2\,{\pi
}{\bar{\sigma}})^{2+\frac{\Gamma}{z}}\,z\,\bar{r}}} \,\left( {\frac
{z\bar{r}}{l}} \right) ^{\frac{4\,p-3}{z (2\,p-1)}} + \frac{m (z+2)}
{ 2 \kappa \left( 2 \pi \bar{\sigma} \right) ^{{\frac
{z+1}{z}}}{l}}, \nonumber
\end{equation*}
yielding to
$$
\Delta_{0}=\frac{z m \Omega_{1}}{2 \kappa l \left(2 \pi
\bar{\sigma}\right)^{\frac{z+1}{z}}}.
$$
The magnetic charge and potential, as before, are purely imaginary
and read
\begin{eqnarray}
{\cal Q}_{m}&=& \frac{i \,p\,\sqrt{2}}{2\,\kappa}\, \left( {\frac
{z}{l}} \right) ^{{ \frac {2\,p-1}{zp}}} \,{\bar{\Sigma}}^{{\frac
{2\,p-1}{2 p}}}\,{\mu}^{{\frac { \left( z-1 \right) \left( 2\,p-1
\right) }{p\,\sqrt {2\,(z-1)}}}}\Omega_{1}
, \\\nonumber\\
\Phi_{m}&=&{\frac {i \,\sqrt {2}\,\left( 2\,p-1 \right) \,z} {[
2\left(z-2 \right) p +3-z]}} \left( {\frac {z}{l}} \right) ^{{\frac
{p-1}{ \left( 2\,p-1 \right) zp}}}
{\bar{\Sigma}}^{\frac{1}{2\,{p}}}\,{\mu}^ {-{\frac { \left( z-1
\right)  \left( 2\,p-1 \right) }{p\,\sqrt {2(z-1)}}}}
\nonumber\\
&\times&\left( {\frac {l}{2 \pi \,\bar{\sigma}\,z}} \right)
^{-{\frac {[ 2\left( z-2 \right) p+3-z]}{ \left( 2\,p-1 \right)
z}}}.
\end{eqnarray}
As a matter of check, one can see that all the expressions involve
in the nonlinear case reduce to those obtained in the previous
sub-section in the linear limiting case $p=1$.

Finally, it is straightforward to check that the formula
(\ref{cardycharged}) with $\alpha$ given by (\ref{alphap}) fits
perfectly with the Wald formula (\ref{Wald1nonlinear}).

\subsection{Case of Einstein gravity with a Proca and Maxwell fields}
We now consider the case of Einstein gravity with a Proca field
$A_{(1)_{\mu}}$ together with a Maxwell field $A_{(2)_{\mu}}$ whose
Lagrangian is given by
\begin{eqnarray}
{\cal L}=&&R-2\Lambda
-\frac{1}{4}F_{(1)\alpha\beta}F_{(1)}^{\,\,\alpha\beta}
-\frac{1}{2}m^2A_{(1)\alpha}A_{(1)}^{\alpha}\nonumber\\
&&-\frac{1}{4}F_{(2)\alpha\beta}F_{(2)}^{\,\,\alpha\beta}
\label{action}
\end{eqnarray}
with
$F_{(i)\alpha\beta}=\partial_{\alpha}A_{(i)\beta}-\partial_{\beta}A_{(i)\alpha}$
for $i=1, 2$.

In this case, the electrically charged Lifshitz black hole solution
exists only for $z=2$; the metric function (\ref{lifbhsoln2}) and
the Proca and Maxwell fields read \cite{Pang:2009pd}
\begin{eqnarray}
\label{functions1}
&&f(r)=1-\left(\frac{r_h}{r}\right)^{2},\, A_{(1)t}(r)=\left(\frac{r}{l}\right)^{2}f(r),\nonumber\\
&& \label{functions2} F_{(2)tr}(r)=\frac{\sqrt{2}}{l^2}\,r_{h},
\end{eqnarray}
while the parameters must be fixed as follows
\begin{eqnarray*}
m=\frac{\sqrt{2}}{l},\qquad \Lambda=-\frac{5}{2l^2}.
\end{eqnarray*}
For this solution, the Wald entropy $\mathcal{S}_{\mathrm{W}}$ and
the Hawking temperature are given by
\begin{align}\label{entropytemsoln}
\mathcal{S}_{\mathrm{W}}&= {\frac {4\,\pi\,r_h\,\Omega_{1}}{l}},
\quad T=\frac{r_{h}^2}{2\,\pi\,l^{3}}.
\end{align}
The expressions of the surface term and Noether potential
(\ref{eq:theta}-\ref{eq:K}) read
\begin{align}
\int_{0}^{1}\!\!\!\!ds\,\Theta^{r}={}& {\frac
{2\,r\,r_h^2}{{l}^{4}}}, \qquad \delta K^{rt}=-{\frac
{2\,r\,r_h^2}{{l}^{4}}},
\end{align}
which in turn implies that the mass $\Delta=0$. This solution with
vanishing mass can be interpreted as an extremal charged Lifshitz
black hole as it occurs for examples in Refs.
\cite{Liu:2014dva,Bravo-Gaete:2015xea}. Nevertheless, the electric
charge ${\cal Q}_e$ and the electric potential $\Phi_e$ are
non-vanishing and given by
\begin{eqnarray}
{\cal
Q}_e=\frac{\sqrt{2}\,\Omega_1}{l}\left(\frac{r_h}{l}\right),\quad
\Phi_e=-\sqrt{2}\left(\frac{r_h}{l}\right)^2.
\end{eqnarray}
It is easy to verify that the first law of thermodynamics holds
\begin{eqnarray}
d{\Delta}=0=Td\mathcal{S}_{\mathrm{W}}+\Phi_e d{\cal Q}_e,
\label{ftlwa}
\end{eqnarray}
and the Smarr formula (\ref{smarrF1charged}) reads in this case
\begin{eqnarray}
\Delta=0=\frac{1}{3}\left(T\mathcal{S}_{\mathrm{W}}+\Phi_e {\cal
Q}_e\right),
\end{eqnarray}
that is the constant $\alpha$ appearing in the generic formula
(\ref{smarrF1charged}) is $\alpha=\frac{1}{3}$.

The corresponding soliton is given by the line element
(\ref{lifsolsoln1}) with $z=2$ where the metric function and the
gauge fields are given by
\begin{eqnarray}
&& h(\bar{r})=1-\frac{l}{2\bar{r}},\nonumber\\
&&
A_{(1)\bar{\varphi}}=2i\left(\frac{\bar{r}}{l}\right)\,h(\bar{r}),\quad
F_{(2)\bar{r}\bar{\varphi}}=i\left(l\bar{r}\right)^{-1/2}.
\end{eqnarray}
Along the same lines as before, the  Noether potential together with
the surface term take the following forms
\begin{align}
\int_{0}^{1}\!\!\!\!\,d{s}\,\Theta^{\bar{r}}=\frac{2}{l}\, \left(
{\frac {2\bar{r}}{l}} \right) ^{\frac{1}{2}},\qquad \Delta
K^{\bar{r}\bar{t}}=-\frac{2}{l}\, \left( {\frac {2\bar{r}}{l}}
\right) ^{\frac{1}{2}},
\end{align}
and as in the electric case, the mass of the soliton is vanishing
$\Delta_0=0$. The magnetic charge and potential are purely imaginary
and read
\begin{eqnarray}
{\cal Q}_m=i\frac{\sqrt{2}}{l}\Omega_1,\qquad \Phi_m=-i\sqrt{2},
\end{eqnarray}
and it is a matter of check that the formula (\ref{cardycharged})
with $z=2$ and $\alpha=1/3$ fits perfectly with the Wald formula
(\ref{entropytemsoln}).

\section{Generalization for charged Lifshitz black holes with
hyperscaling violation}
In the anisotropic extension of the AdS/CFT correspondence, there
exists another dual metric of interest, the so-called hyperscaling
violation spacetime whose line element can be parameterized as
follows
\begin{eqnarray}
ds^{2}&=&\frac{1}{r^{\frac{{2\theta}}{D-2}}}\Big[-r^{2z}
{dt^2}+\frac{dr^2}{r^2 }+r^2d\vec{x}^2\Big]. \label{HSV}
\end{eqnarray}

In this case, the anisotropic transformations (\ref{anisyymetry})
together with $r\to \lambda^{-1} r$ act rather like a conformal
transformation, $ds^2\to \lambda^{2\theta/(D-2)}ds^2$. Note also
that this metric reduces to the Lifshitz metric (\ref{Lifmetric}) in
the limiting case $\theta=0$.

In Refs. \cite{Shaghoulian:2015dwa,Bravo-Gaete:2015wua}, it was
shown that if the entropy $S$ scales with respect to the temperature
$T$ as
\begin{eqnarray} S\sim T^{\frac{d_{\tiny{\mbox{eff}}}}{z}},
\label{scaleST}
\end{eqnarray}
where $d_{\tiny{\mbox{eff}}}$ is the effective spatial
dimensionality, and where $z$ is the dynamical exponent, the formula
(\ref{cardyuncharged}) in the uncharged case becomes
\begin{eqnarray}
{\cal S}=\frac{2\pi}{d_{\tiny{\mbox{eff}}}}
(z+d_{\tiny{\mbox{eff}}})\left(\frac{\Delta_0\,
d_{\tiny{\mbox{eff}}}}{z}\right)^{\frac{z}{z+d_{\tiny{\mbox{eff}}}}}\,\Delta^{\frac{d_{\tiny{\mbox{eff}}}}{z+d_{\tiny{\mbox{eff}}}}}.
\label{cardyunchargedHVM}
\end{eqnarray}
Repeating the same exercise than in the Lifshitz case, the first law
$d\Delta=Td{\cal S}$ allows to express the mass as
\begin{eqnarray}
\Delta=\left({2\pi
T}\right)^{\frac{z+d_{\tiny{\mbox{eff}}}}{z}}\left(\frac{\Delta_0\,
d_{\tiny{\mbox{eff}}}}{z}\right), \label{massHvm}
\end{eqnarray}
and the Smarr formula becomes
\begin{eqnarray}
\Delta=\frac{d_{\tiny{\mbox{eff}}}}{z+d_{\tiny{\mbox{eff}}}}\,T\,{\cal
S}. \label{SmarrHvm}
\end{eqnarray}
We may note that the expressions
(\ref{scaleST}-\ref{cardyunchargedHVM}-\ref{massHvm}-\ref{SmarrHvm})
with $d_{\tiny{\mbox{eff}}}=1$ reduce to those obtained in the
Lifshitz case.

Now by a certain analogy with the charged Lifshitz case, the Cardy
formula for electrically charged black holes with hyperscaling
violation should be
\begin{widetext}
\begin{eqnarray}
{\cal
S}=\frac{2\pi}{d_{\tiny{\mbox{eff}}}}(z+d_{\tiny{\mbox{eff}}})\Big(\vert
\frac{\Delta_0}{z}d_{\tiny{\mbox{eff}}}+\alpha\, \Phi_m {\cal Q}_m
\vert^z\,\vert\Delta-\alpha\, \Phi_e {\cal Q}_e
\vert^{d_{\tiny{\mbox{eff}}}}
\Big)^{\frac{1}{z+d_{\tiny{\mbox{eff}}}}}. \label{cardychargedHvm}
\end{eqnarray}
\end{widetext}
As before, the constant $\alpha$ is the one appearing in the charged
version of the Smarr formula in the hyperscaling case, namely
\begin{eqnarray}
\Delta=\frac{d_{\tiny{\mbox{eff}}}}{z+d_{\tiny{\mbox{eff}}}}\,T\,{\cal
S}+\alpha\, \Phi_e {\cal Q}_e. \label{SmarrHvmcharged}
\end{eqnarray}

Let us now verify this formula for the charged hyperscaling
violation black hole derived in \cite{Dehghani:2015gza} for which
the Lagrangian is given by
\begin{eqnarray}
{\cal L}=\frac{1}{2\kappa}\left(R-\frac{1}{2}(\partial\phi)^2+V(\phi)-\sum_{i=1}^{2}\frac{1}{4}e^{\lambda_i\phi}F_{(i)\mu\nu}F_{(i)}^{\,\,\mu\nu}\right),\nonumber\\
\end{eqnarray}
where the potential is
$$
V(\phi)=-2\Lambda e^{\gamma\phi}.
$$
The solution as reported in \cite{Dehghani:2015gza}, again after
some redefinitions of the constant, reads
\begin{eqnarray}\label{HVM}
ds^{2}&=&\frac{1}{r^{2\theta}}\,\Big[-r^{2z}\,f(r)
{dt^2}+\frac{dr^2}{r^2 \,f(r) }+r^2d\vec{\varphi}^2\Big],
\end{eqnarray}
where
\begin{subequations}
\begin{eqnarray}
f(r)&=&1-m\left(\frac{r_{h}}{r}\right)^{z+1-\theta}
+(m-1)\left(\frac{r_{h}}{r}\right)^{2z-2\theta}, \label{metricfunctionhvm}\\
\nonumber\\
F_{(1) rt}&=&\sqrt { 2\,\left( z-1 \right)  \left( 1-\theta+z
\right) }\,\,{\mu}^ {{\frac {\sqrt {2}}{2 \sqrt
{(1-\theta)(z-\theta-1)}}}}\nonumber\\
&\times&\,{r}^{z-\theta},\\ \nonumber\\
F_{(2) rt}&=&\sqrt { 2 \left( 1-\theta \right)  \left( z-\theta-1
\right) \left( m-1 \right) }\,\,{\mu}^{-{\frac {\sqrt
{z-\theta-1}\sqrt {2}} {2 \sqrt {1-\theta}}}}
\nonumber\\
&\times& \left(\frac{r_h}{r}\right)^{z-\theta},\\
\nonumber\\
e^{\phi}&=&\mu r^{\sqrt{2(1-\theta)(z-\theta-1)}} \label{phihvm},
\end{eqnarray}
\end{subequations}
while the parameters are fixed as
\begin{eqnarray}
\lambda_1&=&-\sqrt{\frac {2}{(1-\theta)(z-\theta-1)}},\quad
\lambda_2=\sqrt{\frac{2(z-\theta-1)}{1-\theta}},\nonumber\\
\Lambda&=&\frac{1}{2}\, \left( 1-\theta+z \right)  \left(
z-\theta\right) {\mu}^{-{ \frac {\theta\,\sqrt {2}}
{\sqrt {(1-\theta)(z-\theta-1)}}}},\nonumber\\
\gamma&=&{\frac {2\,\theta}{\sqrt {2\,(1-\theta)(z-\theta-1)}}},
\end{eqnarray}
For this solution, the Wald entropy is given by
\begin{eqnarray}\label{waldhvm}
 {\cal S}_{W}&=&{\frac {2 \pi\,{
{r_h}}^{1-\theta}\,\Omega_{1}}{\kappa}},
\end{eqnarray}
while the Hawking temperature is
\begin{eqnarray}
T&=&{\frac { \left[  \left( m-2 \right) \theta-m \left( z-1
 \right) +2\,z \right] {r_h}^{z}}{4 \pi }}=\rho\,{r_h}^{z},
\end{eqnarray}
where for simplicity  we define
\begin{eqnarray}\label{rho}
\rho=\frac {\left( m-2 \right) \theta}{4 \pi }+\sigma,
\end{eqnarray}
with $\sigma$ given in (\ref{expressofA}). In this case, the
electric charge together with the potential read respectively
\begin{align}
{\cal{Q}}_{e}&=\frac{ \sqrt { 2 \left( 1-\theta \right) \left(
z-\theta-1 \right) \left( m-1 \right) }\,{\mu}^{{\frac {\sqrt
{2(z-\theta-1)}}{2 \sqrt {1-\theta}}}}\,
{{r_h}}^{z-\theta}\,\Omega_{1}
}{2 \kappa},\\
\Phi_{e}&= \sqrt{{\frac { 2\,\left( 1-\theta \right)  \left( m-1
\right) }{z-\theta-1}}}\,{\mu}^{-{\frac {\sqrt {2(z-\theta-1)}
}{2\,\sqrt {1-\theta}}}}\,r_{h}.
\end{align}
The variation of the Noether potential together with the surface
term are obtained as
\begin{align}
\delta K^{r t}&=-{\frac { \left( z-\theta-1 \right)\,
m\,{r_h}^{1-\theta+z}} {2\,
\kappa}}\nonumber\\
&-{\frac { \left(m-1 \right) \left( \theta-1 \right) \,{{r_h}}^{2\,z
-2\,\theta}}{\kappa}}\,{r}^{\theta-z+1}
,\nonumber\\
\int_{0}^{1}\!\!\!\!d s\Theta^{r}&= {\frac { \left(z -2\,\theta
\right)\, m\,\,{r_h}^{1-\theta+z}}{ 2\,\kappa}} \nonumber\\
&+{\frac { \left(m-1 \right) \left( \theta-1 \right) \,{{r_h}}^{2\,z
-2\,\theta}}{\kappa}}\,{r}^{\theta-z+1},
\end{align}
yielding to the same expression of the mass as the one found in
\cite{Dehghani:2015gza}, namely
\begin{equation}\label{masshvbh}
\Delta=\frac{m\,(1-\theta)\,\Omega_{1}} {2
\kappa}\,{r_h}^{1-\theta+z}.
\end{equation}
From all these expressions, one can easily check the validity of the
first law while the effective spatial dimensionality is given by
\begin{eqnarray}
d_{\tiny{\mbox{eff}}}=1-\theta,
\end{eqnarray}
and the Smarr formula (\ref{SmarrHvmcharged}) is realized with a
constant $\alpha$ chosen as
\begin{eqnarray}
\alpha=\frac{z-\theta}{z+1-\theta}. \label{alphaHVM}
\end{eqnarray}

On the other hand, the soliton counterpart for the hyperscaling
violation metric (\ref{HVM}) with the metric function
(\ref{metricfunctionhvm}), obtained through a double Wick rotation,
has the following form
\begin{eqnarray}\label{HVMsol}
ds^{2}&=&\frac{1}{r^{2\theta}}\,\Big[-r^{2z} {dt^2} +\frac{d
r^2}{r^2 \,h(r) }+r^2\,h(r) d\vec{\varphi}^2\Big],
\end{eqnarray}
where the metric function $h$ is defined as
\begin{eqnarray}
h(r)=1-\frac{m} { \left( 2 \pi\rho \right) ^{{\frac
{1+z-\theta}{z}}} {r}^{1+z-\theta}}+\frac{\left( m-1 \right)} {
\left( 2 \pi \rho\right) ^{{\frac {2\, z-2\,\theta}{z}}}
{r}^{2\,z-2\,\theta}},\nonumber\\
\end{eqnarray}
with $\rho$ being given by Eq. (\ref{rho}). The gauge fields read in
this case
\begin{eqnarray}
F_{(1) r \varphi}&=&i \sqrt { 2\,\left( z-1 \right)  \left(
1-\theta+z \right) }\,{\mu}^ {{\frac {\sqrt {2}}{2 \sqrt
{(1-\theta)(z-\theta-1)}}}}\,{r}^{z-\theta},\nonumber\\
\\
F_{(2) r \varphi}&=&-i\, \sqrt{\frac{2 \left( 1-\theta \right)
\left( m-1 \right) }{\left( z-\theta-1 \right)}}\,{{\mu}^{-{\frac
{\sqrt {2\,(z-\theta-1)}} {2 \sqrt{1-\theta}}}}}\nonumber\\
&\times&\left(\frac{1}{2 \pi \rho}\right)^{\frac{z-\theta}{z}}\,
r^{-z+\theta+1},
\end{eqnarray}
while the dilaton is given by Eq. (\ref{phihvm}).

As before, choosing the Killing vector $\xi^{t}=(1,0,0)$, the
variation of the Noether potential and the surface term are
calculated as
\begin{align*}
\delta K^{r t}&=-\frac{(\theta-1)\,(m-1)} {\kappa\,(2 \pi
\rho)^{\frac{2z-2\theta}{z}}\,r^{z-\theta-1}} + \frac{
\left(\theta-1 \right) m }{{\kappa}\,  \left( 2\,\pi\,\rho \right)
^{{ \frac {1-\theta+z}{z}}}}
,\nonumber\\
\int_{0}^{1}\!\!\!\!d s\Theta^{r}&=\frac{(\theta-1)\,(m-1)}
{\kappa\,(2 \pi \rho)^{\frac{2z-2\theta}{z}}\, r^{z-\theta-1}} +
\frac{ \left(z+2-2\,\theta \right) m }{2 {\kappa}\,\left(
2\,\pi\,\rho  \right) ^{{ \frac {1-\theta+z}{z}}}},
\end{align*}
which, in turn, implies that
\begin{eqnarray}
\Delta_{0}= \frac{z\,m\, \Omega_{1}}{ 2 \kappa \left( 2 \pi \rho
\right)^{\frac{1+z-\theta}{z}}}.
\end{eqnarray}
Finally, the magnetic charge and the magnetic potential read
\begin{eqnarray}
{\cal{Q}}_{m}&=& \frac{i \sqrt {2 (1-\theta)(z-\theta-1)
(m-1)}\,{\mu}^ { {\frac {\sqrt {2\,(z-\theta-1)}}{2 \sqrt {1-\theta
}}}}\,\Omega_{1}}{2 \kappa \left( 2 \pi \rho\, \right)
^{\frac{z-\theta}{z}}},\nonumber
\\
\Phi_{m}&=&i \sqrt{\frac{2 (1-\theta)(m-1)}{z-\theta-1}}\, {\mu}^ {
{-\frac {\sqrt {2\,(z-\theta-1)}}{2 \sqrt {1-\theta }}}}\,
\left(\frac{1}{2 \pi \rho}\right)^{\frac{1}{z}}.
\end{eqnarray}
Once again, it is easy to verify that the formula
(\ref{cardychargedHvm}) with the parameter $\alpha$ given by
(\ref{alphaHVM}) correctly fits with the Wald entropy defined in
(\ref{waldhvm}).

\section{The case of AdS charged black holes}
We now consider the isotropic AdS case which corresponds to a
dynamical exponent $z=1$. There exist examples of electrically
charged AdS black holes in three dimensions, and the most popular
one  is the charged BTZ  solution \cite{Banados:1992wn}.
Unfortunately in this case, because of the logarithmic behavior of
the Maxwell electric gauge field, there is not such a  Smarr formula
(\ref{smarrF1charged}) encoding the charged BTZ solution. As a
direct consequence, the Cardy formula given in (\ref{cardycharged})
is not longer valid in such situation. Nevertheless, as shown in
Refs. \cite{Cataldo:2000we,Cardenas:2014kaa}, considering instead a
nonlinear version of the Maxwell action (as the one used in Sec
II.B) and eventually a scalar field nonminimally and conformally
coupled, there exist electrically AdS charged black hole such that
the  electric gauge field $A_t(r)$ exhibits a Coulombian behavior,
that is $A_t(r)\sim r^{-1}$. In what follows, we shall consider a
such particular solution that satisfies a Smarr relation
(\ref{smarrF1charged}) and show again that the Cardy formula
(\ref{cardycharged}) will reproduce the correct value of the
entropy.

We deal with the Lagrangian reported in \cite{Cardenas:2014kaa}
\begin{eqnarray}
{\cal L}=&&\frac{R+2l^{-2}}{2\kappa}-\frac{1}{2}(\partial\phi)^2-\frac{1}{16}R\phi^2-\lambda\phi^6\\
&&+\sigma(-F_{\mu\nu}F^{\mu\nu})^{3/4},\nonumber \label{CFM}
\end{eqnarray}
where the matter part of the action (the scalar field and the
nonlinear electromagnetic action) is chosen such that it enjoys the
conformal invariance. The solution we consider for testing the Cardy
formula   (\ref{cardycharged})  is given by the simplest one found
in \cite{Cardenas:2014kaa},
\begin{eqnarray}
&&f(r)=1+\frac{24\lambda b^2 l^2}{r^2},\,\,
\phi(r)=\sqrt{\frac{b}{r}},\, \quad F_{rt}=\frac{q}{r^{2}},
\end{eqnarray}
where $f(r)$ is the metric function of the line element
(\ref{lifbhsoln2}) with $z=1$, and $\vert q\vert^{3/2}=-\lambda
b^3$. The constant $b$ is strictly positive
 while $\lambda$ is negative.  In this particular case, the quantities of interest read \cite{Cardenas:2014kaa}
\begin{eqnarray}
&&\Delta=\frac{\pi r_h^2}{\kappa l^2},\quad T=\frac{r_h}{2\pi l^2},\quad{\cal S}_{W}=\left(1-\frac{\kappa}{8l\sqrt{-24\lambda}}\right)\frac{4\pi^2 r_h}{\kappa},\nonumber\\
&& {\cal Q}_e=\frac{6\pi
(-\lambda)^{1/3}}{\sqrt{-24\lambda}\,l}r_h,\quad
\Phi_e=\frac{r_h}{24l^2(-\lambda)^{1/3}}, \label{CFMq}
\end{eqnarray}
where the location of the horizon $r_h$ is defined by
$r_h^2=-24\lambda b^2 l^2$, and for simplicity we have assumed that
$q>0$. Having in hands all these quantities, one easily verify that
a Smarr relation (\ref{smarrF1charged}) is satisfied with
$\alpha=1/2$.

The corresponding soliton solution is described by
\begin{eqnarray} &&g(\bar{r})=1-\frac{l^2}{\bar{r}^2},\quad
\phi(\bar{r})=\left(\frac{1}{24\,(-\lambda)}\right)^{1/4}\,\sqrt{\frac{1}{\bar{r}}},
\nonumber \\
&&F_{\bar{r}\bar{\varphi}}(\bar{r})=\frac{i}{24 \,
(-\lambda)^{1/3}\, \bar{r}^{2}},
\end{eqnarray}
where $g(\bar{r})$ is the metric function of (\ref{lifsolsoln1})
with $z=1$. We may note that, as said before, the double Wick
rotation does not yield to a complex scalar field for the soliton
solution.

Along the same lines as before, the surface term together with the
variation of the Noether potential are given by
\begin{align}
&\int_{0}^{1}\!\!\!\!\,d{s}\,\Theta^{\bar{r}}= {\frac {\sqrt
{6}\,\bar{r}}{48\,{l}^{3}\sqrt {-\lambda}}}- { \frac {\sqrt {6}}{48
l\sqrt {-\lambda}\bar{r}}}+{\frac {3}{2\kappa l}} ,\nonumber\\
&\delta K^{\bar{r}\bar{t}}=-{\frac {\sqrt
{6}\,\bar{r}}{48\,{l}^{3}\sqrt {-\lambda}}}+ { \frac {\sqrt {6}}{48
l\sqrt {-\lambda}\bar{r}}}-{\frac {1}{\kappa l}},
\end{align}
yielding to
$$\Delta_0=\frac{1}{2\kappa\,l}\Omega_1=\frac{\pi}{\kappa}.$$
Finally, the magnetic charge and magnetic potential read
\begin{eqnarray}
&&{\cal{Q}}_{m}={\frac {i\sqrt {6}\,\pi }{2\,
({-\lambda})^{1/6}}},\qquad \Phi_{m}={\frac {i}{24\,l\,
({-\lambda})^{1/3}}}.
\end{eqnarray}
Hence, as in the anisotropic case, the charged version of the Cardy
formula (\ref{cardycharged}) with $\alpha=1/2$ and $z=1$ gives the
correct value of the entropy (\ref{CFMq}).

\section{Concluding remarks}
Our starting point was the observation that in the case of Lifshitz
black holes whose only charge is the mass, the general asymptotic
formula for the asymptotic growth of number of states derived in
\cite{Gonzalez:2011nz} naturally implies the emergence of a Smarr
formula given by (\ref{smarrF1}) in $D=3$. In our search of
generalizing the Cardy formula to the case of electrically charged
Lifshitz black holes, we have proposed a formula compatible with a
charged version of the Smarr formula of the form
(\ref{smarrF1charged}). We have tested the viability of this formula
in three different examples where charged Lifshitz black holes
obeying a Smarr relation were known. We have extended our analysis
to the other class of charged black hole solutions with anisotropic
symmetry, namely those exhibiting a hyperscaling violation.

In the case of the isotropic charged AdS black holes, we have shown
that the absence of a Smarr relation for the charged BTZ solution
renders our formula (\ref{cardycharged}) inappropriate. The absence
of a Smarr relation of the form (\ref{smarrF1charged}) is mainly due
to the logarithmic behavior of the Maxwell gauge field. It seems
that in this case, the appropriate formula should be the
Cardy-Verlinde formula  \cite{Verlinde:2000wg} where the Smarr
relation is augmented by a pressure term, see \cite{Setare:2009hw}.

Nevertheless, replacing the standard Maxwell theory by its nonlinear
and conformal generalization, asymptotically charged AdS black holes
are known with a gauge field behaving as a Coulomb one. In a simple
example of such solution giving in Ref. \cite{Cardenas:2014kaa}, we
have again tested the viability of the Cardy formula after ensuring
that this Coulombian solution was as well satisfying a Smarr
relation.

As a natural extension of this work, it will be desirable to test
this formula in much more examples, and particulary to those
involving higher-order gravity theories in three dimensions. This
task can be interesting by itself in the hyperscaling violation
case, since as shown in \cite{Bravo-Gaete:2015wua}, the spatial
effective dimensionality $d_{\tiny{\mbox{eff}}}$ may vary in
function of the order of the gravity theories involved in the
action.

Also there exists a generalization of the Smarr relation in the case
of AdS black holes for which the cosmological constant is viewed as
a dynamical variable. In a very recent paper, the authors of Ref.
\cite{Karch:2015rpa} showed that such generalization of the Smarr
relation can be understood from a dual holographic point of view.
Extension to the Lifshitz case can also be an interesting work to
deal with.

\begin{acknowledgments}
This work is partially supported by grant 1130423 from FONDECYT and
from CONICYT with grant 21130136. This project is also partially
funded by Proyectos CONICYT- Research Council UK - RCUK
-DPI20140053.
\end{acknowledgments}




\begin{thebibliography}{99}



\bibitem{Maldacena:1997re}
  J.~M.~Maldacena,
  Adv.\ Theor.\ Math.\ Phys.\  {\bf 2}, 231 (1998)
  [hep-th/9711200].


\bibitem{Kachru} S. Kachru, X. Liu and M. Mulligan, Phys.
    Rev. \textbf{D78} (2008) 106005, [arXiv:0808.1725].


\bibitem{Azeyanagi:2009pr}
  T.~Azeyanagi, W.~Li and T.~Takayanagi,
  JHEP {\bf 0906}, 084 (2009)
  [arXiv:0905.0688 [hep-th]].





\bibitem{Correa:2014ika}
  F.~Correa, M.~Hassaine and J.~Oliva,
  Phys.\ Rev.\ D {\bf 89}, 124005 (2014)
  [arXiv:1403.6479 [hep-th]].


\bibitem{Ayon-Beato:2015jga}
  E.~Ayón-Beato, M.~Bravo-Gaete, F.~Correa, M.~Hassaïne, M.~M.~Juárez-Aubry and J.~Oliva,
  Phys.\ Rev.\ D {\bf 91}, no. 6, 064006 (2015)
  [arXiv:1501.01244 [gr-qc]].

\bibitem{Bravo-Gaete:2013dca}
  M.~Bravo-Gaete and M.~Hassaine,
  Phys.\ Rev.\ D {\bf 89}, no. 10, 104028 (2014)
  [arXiv:1312.7736 [hep-th]].




\bibitem{AyonBeato:2009nh}
  E.~Ayon-Beato, A.~Garbarz, G.~Giribet and M.~Hassaine,
  Phys.\ Rev.\ D {\bf 80}, 104029 (2009)
  [arXiv:0909.1347 [hep-th]].

\bibitem{AyonBeato:2010tm}
  E.~Ayon-Beato, A.~Garbarz, G.~Giribet and M.~Hassaine,
  JHEP {\bf 1004}, 030 (2010)
  [arXiv:1001.2361 [hep-th]].
\bibitem{Lu:2012xu}
  H.~Lu, Y.~Pang, C.~N.~Pope and J.~F.~Vazquez-Poritz,
  Phys.\ Rev.\ D {\bf 86}, 044011 (2012)
  [arXiv:1204.1062 [hep-th]].


\bibitem{Giacomini:2012hg}
  A.~Giacomini, G.~Giribet, M.~Leston, J.~Oliva and S.~Ray,
  Phys.\ Rev.\ D {\bf 85}, 124001 (2012)
  [arXiv:1203.0582 [hep-th]].

\bibitem{Oliva:2012zs}
  J.~Oliva and S.~Ray,
  Phys.\ Rev.\ D {\bf 86}, 084014 (2012)
  [arXiv:1201.5601 [gr-qc]].


\bibitem{Liu:2014dva}
  H.~S.~Liu and H.~Lu,
  JHEP {\bf 1412}, 071 (2014)
  [arXiv:1410.6181 [hep-th]].




\bibitem{Devecioglu:2011yi}
  D.~O.~Devecioglu and O.~Sarioglu,
  Phys.\ Rev.\ D {\bf 83}, 124041 (2011)
  [arXiv:1103.1993 [hep-th]].


\bibitem{Bravo-Gaete:2015xea}
  M.~Bravo-Gaete and M.~Hassaine,
  Phys.\ Rev.\ D {\bf 91}, no. 6, 064038 (2015)
  [arXiv:1501.03348 [hep-th]].


\bibitem{Smarr}
L. Smarr, Phys.\ Rev.\ Lett.\ {\bf 30}, 71 (1973) [Erratum:  Phys.\
Rev.\ Lett.\ {\bf 30}, 521 (1973)].


\bibitem{Brenna:2015pqa}
  W.~G.~Brenna, R.~B.~Mann and M.~Park,
  Phys.\ Rev.\ D {\bf 92}, no. 4, 044015 (2015)
  [arXiv:1505.06331 [hep-th]].


\bibitem{Gonzalez:2011nz}
  H.~A.~Gonzalez, D.~Tempo and R.~Troncoso,
  JHEP {\bf 1111}, 066 (2011)
  [arXiv:1107.3647 [hep-th]].





\bibitem{Taylor} M. Taylor, ``Non-relativistic
    holography,'' [arXiv:0812.0530].

\bibitem{DanielssonI} U. Danielsson and L. Thorlacius, JHEP \textbf{0903} (2009) 070,
    [arXiv:0812.5088].

\bibitem{Pang:2009pd}
  D.~W.~Pang,
  JHEP {\bf 1001}, 116 (2010)
  [arXiv:0911.2777 [hep-th]].

\bibitem{Alvarez:2014pra}
  A.~Alvarez, E.~Ayon-Beato, H.~A.~Gonzalez and M.~Hassaine,
  JHEP {\bf 1406}, 041 (2014)
  [arXiv:1403.5985 [gr-qc]].



\bibitem{Zangeneh:2015uwa}
  M.~K.~Zangeneh, A.~Sheykhi and M.~H.~Dehghani,
  Phys.\ Rev.\ D {\bf 92}, no. 2, 024050 (2015)
  [arXiv:1506.01784 [gr-qc]].

\bibitem{Dehghani:2013mba}
  M.~H.~Dehghani, C.~Shakuri and M.~H.~Vahidinia,
  Phys.\ Rev.\ D {\bf 87}, no. 8, 084013 (2013)
  [arXiv:1306.4501 [hep-th]].







\bibitem{Tarrio:2011de}
  J.~Tarrio and S.~Vandoren,
  JHEP {\bf 1109}, 017 (2011)
  [arXiv:1105.6335 [hep-th]].




\bibitem{Kim:2013zha}
  W.~Kim, S.~Kulkarni and S.~-H.~Yi,
  Phys.\ Rev.\ Lett.\  {\bf 111}, no. 8, 081101 (2013)
  [arXiv:1306.2138 [hep-th]].






\bibitem{Gim:2014nba}
  Y.~Gim, W.~Kim and S.~H.~Yi,
  JHEP {\bf 1407}, 002 (2014)
  [arXiv:1403.4704 [hep-th]].



\bibitem{nonlinMax}
  M.~Hassaine and C.~Martinez,
  Phys.\ Rev.\ D {\bf 75}, 027502 (2007), [hep-th/0701058];  M.~Hassaine and C.~Martinez,
  Class.\ Quant.\ Grav.\  {\bf 25}, 195023 (2008)
  [arXiv:0803.2946 [hep-th]];  S.~H.~Hendi, S.~Panahiyan and H.~Mohammadpour,
  Eur.\ Phys.\ J.\ C {\bf 72}, 2184 (2012)
  [arXiv:1501.05841 [gr-qc]]; S.~H.~Hendi,
  Adv.\ High Energy Phys.\  {\bf 2014}, 697863 (2014)
  [arXiv:1405.6997 [physics.gen-ph]].



\bibitem{Shaghoulian:2015dwa}
  E.~Shaghoulian,
  arXiv:1504.02094 [hep-th].



\bibitem{Bravo-Gaete:2015wua}
  M.~Bravo-Gaete, S.~Gomez and M.~Hassaine,
  Phys.\ Rev.\ D {\bf 91}, no. 12, 124038 (2015)
  [arXiv:1505.00702 [hep-th]].















\bibitem{Dehghani:2015gza}
  M.~H.~Dehghani, A.~Sheykhi and S.~E.~Sadati,
  Phys.\ Rev.\ D {\bf 91}, no. 12, 124073 (2015)
  [arXiv:1505.01134 [hep-th]].

\bibitem{Banados:1992wn}
  M.~Banados, C.~Teitelboim and J.~Zanelli, Phys.\ Rev.\ Lett.\  {\bf 69}, 1849 (1992)
  [hep-th/9204099].

 \bibitem{Cataldo:2000we}
  M.~Cataldo, N.~Cruz, S.~del Campo and A.~Garcia,
  Phys.\ Lett.\ B {\bf 484}, 154 (2000)
  [hep-th/0008138].


\bibitem{Cardenas:2014kaa}
  M.~Cardenas, O.~Fuentealba and C.~Martinez,
  Phys.\ Rev.\ D {\bf 90}, no. 12, 124072 (2014)
  [arXiv:1408.1401 [hep-th]].


\bibitem{Verlinde:2000wg}
  E.~P.~Verlinde, ``On the holographic principle in a radiation dominated universe,''
  hep-th/0008140.


\bibitem{Setare:2009hw}
  M.~R.~Setare and M.~Jamil,
  Phys.\ Lett.\ B {\bf 681}, 469 (2009)
  [arXiv:0912.0861 [hep-th]].


\bibitem{Karch:2015rpa}
  A.~Karch and B.~Robinson, ``Holographic Black Hole Chemistry,
  arXiv:1510.02472 [hep-th].


\end{thebibliography}
\end{document}